\PassOptionsToPackage{table}{xcolor}
\documentclass[sigconf,screen]{acmart}

\copyrightyear{2024}
\acmYear{2024}
\setcopyright{acmlicensed}
\acmConference[ISSTA '24]{Proceedings of the 33rd ACM SIGSOFT International Symposium on Software Testing and Analysis}{September 16--20, 2024}{Vienna, Austria}
\acmBooktitle{Proceedings of the 33rd ACM SIGSOFT International Symposium on Software Testing and Analysis (ISSTA '24), September 16--20, 2024, Vienna, Austria}
\acmDOI{10.1145/3650212.3680368}
\acmISBN{979-8-4007-0612-7/24/09}

\AtBeginDocument{%
  \providecommand\BibTeX{{%
    \normalfont B\kern-0.5em{\scshape i\kern-0.25em b}\kern-0.8em\TeX}}}

\usepackage{svg}
\usepackage{color}
\usepackage{float}
\usepackage{overpic} 
\usepackage{makecell}
\usepackage {multirow}
\usepackage{listings}
\usepackage{xcolor}
\usepackage{algpseudocode}
\usepackage{dashrule}
\usepackage{enumitem}
\usepackage{subfigure}
\usepackage{colortbl}
\usepackage{balance} 
\definecolor{tablered}{RGB}{231, 111, 81}
\definecolor{tableblue}{RGB}{70, 130, 180}
\usepackage[linesnumbered,ruled,vlined]{algorithm2e}

\newcommand{\xzp}[1]{\textcolor{black}{{#1}}}

\newboolean{showcomments}
\setboolean{showcomments}{true}
\ifthenelse{\boolean{showcomments}}
 { \newcommand{\mynote}[2]{
      \fbox{\bfseries\sffamily\scriptsize#1}
        {\small$\blacktriangleright$\textsf{\emph{#2}}$\blacktriangleleft$}}}
        { \newcommand{\mynote}[2]{}}

\newcommand{\toolName}{\textsc{\textbf{SelfPiCo}}\xspace}

\begin{document}
\title{\toolName: Self-Guided Partial Code Execution with LLMs}

%
\author{Zhipeng Xue}
\affiliation{%
  \institution{Zhejiang University}
  \city{Hang Zhou}
  \country{China}
}
\email{zhipengxue@zju.edu.cn}

\author{Zhipeng Gao}
\authornote{This is the corresponding author}
\affiliation{%
  \institution{Shanghai Institute for Advanced Study - Zhejiang University}
    \city{Shang Hai}
  \country{China}
}
\email{zhipeng.gao@zju.edu.cn}

\author{Shaohua Wang}
\affiliation{%
  \institution{Central University of Finance and Economics}
      \city{Bei Jing}
  \country{China}
}
\email{davidshwang@ieee.org}

\author{Xing Hu}
\affiliation{%
  \institution{Zhejiang University}
    \city{Hang Zhou}
  \country{China}
}
\email{xinghu@zju.edu.cn}

\author{Xin Xia}
\affiliation{%
  \institution{Zhejiang University}
    \city{Hang Zhou}
  \country{China}
}
\email{xin.xia@acm.org}

\author{Shanping Li}
\affiliation{%
  \institution{Zhejiang University}
    \city{Hang Zhou}
  \country{China}
}
\email{shan@zju.edu.cn}

%
\renewcommand{\shortauthors}{Xue et al.}

\begin{abstract}
Code executability plays a vital role in software debugging and testing (e.g., detecting runtime exceptions or assertion violations). 
However, code execution, especially partial or arbitrary code execution, is a non-trivial task due to missing definitions and complex third-party dependencies. 
To make partial code (such as code snippets posted on the web or code fragments deep inside complex software projects) executable, the existing study has proposed a machine learning model to predict the undefined element types and inject the pre-defined dummy values into execution. 
However, the performance of their tool is limited due to its simply designed dummy values and the inability to continue learning. 
In this paper, we design and implement a novel framework, named \toolName (\textbf{\underline{Self}}-Guided \textbf{\underline{P}}art\textbf{\underline{i}}al \textbf{\underline{Co}}de Executor), to dynamically guide partial code execution by incorporating the open-source LLM (i.e., Code Llama) within an interactive loop. 
Particularly, \toolName leverages few-shot in-context learning and chain-of-thought reasoning to elicit human knowledge and logical reasoning based on fine-tuning the Code Llama model.
\toolName continuously learns from code execution results and refines its predictions step after step. 
Our evaluations demonstrate that \toolName can execute 72.7\% and 83.3\% of all lines in the open-source code and Stack Overflow snippets, outperforming the most recent state-of-the-art Lexecutor by 37.9\% and 33.5\%, respectively. 
Moreover, \toolName successfully detected 18 and 33 runtime type error issues by executing the partial code from eight GitHub software projects and 43 Stack Overflow posts, demonstrating the practical usage and potential application of our framework in practice.
\end{abstract}


%
\settopmatter{printacmref=true}

\maketitle

\section{Introduction}
\label{sec:intro}
To share ideas or programming techniques, developers write code snippets to illustrate specific task solutions and/or demonstrate programming concepts in the software development community, such as Stack Overflow or GitHub~\cite{gao2020generating, gao2020technical, wang2024just, gao2021automating,venkatesh2016client,wang2019extracting}. 
These arbitrary code snippets are often written for illustrative purposes and as quick ways to convey solutions, without implementation detail, which are widely used by developers~\cite{gao2023know, mai2024human, gao2021code2que}. 
Despite the wide adoption of code snippets among developers, 75\% of the code snippets can not be directly executed~\cite{Galappaththi2022DoesTA, Firouzi2020OnTU, yadavally2024learning, yadavally2023partial} and reused.  
This is because a significant number of code snippets are partial and incomplete (i.e., missing variable or function definitions, missing third-party dependencies). 
Therefore, executing arbitrary code snippets written by developers is essential for reusing these code snippets immediately and effectively. 

The capability of executing partial code also facilitates diverse applications of dynamic program analysis, such as taint analysis~\cite{Aldrich2022AugurDT, Karim2020PlatformIndependentDT, Clause2007DytanAG, Sen2013JalangiAS}, vulnerability and bug detection~\cite{Jiang2022ContextSensitiveAD, Fonseca2011FindingCC, Chen2020MUZZTG, li2024empirically, wen2023less, li2023commit, wang2023deepvd, li2021vulnerability,yang2023does,li2019improving}, type inference~\cite{Miyazaki2018DynamicTI, Hattori2020SemistaticTS, Nicolay2013DeterminingDC,li2023deminify}. 
Dynamic analysis provides valuable insights into a program's runtime behavior, capturing information such as actual data inputs, execution traces, and system reactions. 
It has proven to be effective in unveiling various runtime bugs (e.g., memory leaks, buffer overflow, race conditions~\cite{Wang2023ACS, Legunsen2019HowEA}).   
However, for large-scale software projects, it is difficult, if not possible, to run the dynamic analysis tools on any arbitrary code area that is deep inside the project. 
Executing arbitrary code fragment enable us to run dynamic analysis tools on the key components and vulnerable code area (e.g., the newly updated code), without worrying about the complex building procedure and sophisticated third-party dependencies.



To achieve the goal of executing arbitrary code snippets, Souza et al.~\cite{Souza2023LExecutorLE} first proposed Lexecutor, a neural network guided tool to predict and inject missing values into program execution. 
In particular, when a missing element (e.g., variable, attributes, or function calls) is encountered, their approach queries a machine learning model (i.e., CodeT5\cite{Wang2021CodeT5IU}) to predict the element type and inject a pre-defined dummy value instead. 
However, the performance of Lexecutor is still relatively suboptimal in terms of the code coverage on open-source project functions (50.6\%) and Stack Overflow code snippets (61.0\%). 
After empirically investigating their experimental results, two main challenges are observed regarding their approach: (i) the pre-defined dummy values are too simple and inflexible to cover the practical scenarios in the real development environment. 
For instance in Listing~\ref{listing:introlist}, Lexecutor successfully predicts the correct type of \texttt{filter\_cached}, i.e., \textit{Callable}. Then Lexecutor will inject a pre-defined \textit{DummyCall} for it. However, the program will crash during the execution, since the expected return of \texttt{filter\_cached} includes two values, while the pre-defined \textit{DummyCall} returns only a single value.
(ii) the disability of interactive learning. The Lexecutor uses a machine learning model to predict the missing value types, the prediction results are constant when the input samples are fixed. 
It cannot continue learning from the program execution results, which can provide valuable information to guide the model to make more accurate predictions. 
A skilled developer can gain insights from failed execution results to refine predictions step by step. According to the error message in Listing~\ref{listing:introlist}, the skilled developer would rectify the \textit{DummyCall} by returning either two values or an iterable object, e.g., \textit{Tuple}.
Thus the key question we ask in this work is: \textit{can we design models that can continuously learn from code execution results and incrementally refine predictions, ultimately enabling non-executable code to become executable.}

\begin{lstlisting}[caption=A Failure Case of Lexecutor, label=listing:introlist, escapeinside=||]
# Original Code: black/src/black/concurrency.py:
sources, cached = |\underline{filter\_cached}|(cache, sources)
# Lexecutor Injection:
filter_cached = DummyCall(*args)
|\color{red}TypeError: cannot unpack non-iterable DummyObject object|
# SelfPiCo Injection:
def filter_cached(*args):
    return (1, 2)
\end{lstlisting}



Inspired by the impressive capacities of LLMs (Large Language Models) for code comprehension and their great potential for interacting with humans~\cite{feng2020codebert, Wang2021CodeT5IU, mai2024human, yan2023closer, dai2024mpcoder, xue2023acwrecommender, ma2023training, ma2024understand}, in this work, we first investigate incorporating LLMs for the task of executing arbitrary code snippets. 
The key idea of this work is \textbf{LLM-in-the-loop}. 
Compared with human-in-the-loop (HITL) which uses human interaction to aid computers in making decisions, we first introduce the concept of LITL (LLM-in-the-loop), where the LLMs are engaged within an interactive loop for generating useful artifacts. In particular, we design and implement a novel LLMs-based framework, named \toolName, to guide partial code execution. 
The \toolName is constructed by following three components:
\begin{itemize}[leftmargin=*]
    \item \textbf{Interactive Value Predictor.} 
The interactive value predictor is the core module of \toolName,  which includes an \textit{interactive value generator} and an \textit{execution value checker}. 
The \textit{interactive value generator} is responsible for generating likely values for the missing elements (e.g., undefined variables, return values, or missing functions). 
The \textit{execution value checker} is responsible for ensuring the validity of the generated values. 
If the generated values provided by \textit{interactive value generator} fail to execute the given arbitrary code snippet, the \textit{execution value checker} will query back the \textit{interactive value generator} with error execution messages for regenerating new likely values. 
\item \textbf{Complementary Type Predictor.} 
This component serves as a complement module to the interactive value predictor, addressing cases where the interactive value predictor exceeds maximum iterations. 
If the value predictor can not predict appropriate values, 
the complementary type predictor predicts the type of missing element and injects the pre-defined dummy value. 
\item \textbf{Runtime Engine.} The runtime engine instruments the partial code with execution hooks, which catch the exceptions during code execution, and inject values from interactive value predictor to guide partial code execution.
\end{itemize}




\xzp{Automated program repair (APR) techniques aim to generate a patch that passes compilation and test execution and recent studies have leveraged LLMs for fixing bugs (e.g., compilation or execution bugs)~\cite{jin2023inferfix, joshi2023repair, xia2023automated, deligiannis2023fixing, li2022dear, li2020improving}.
The goal of APR overlaps to some extent with our partial code execution. However, there are two significant distinctions between them: (i) \textbf{The goal is different.}
APR aims to fully repair programs to pass all tests, while our task seeks to make partial code executable. 
Our work can be regarded as a base model to enable other dynamic analysis tools for checking partial code.  
Notably, our tool can also assist developers in fixing bugs or code errors (e.g., exposing runtime errors during execution), but fixing bugs is not the final goal of this research. 
(ii) \textbf{The way of interacting with code is different.} APR generates correct patches to fix bugs in buggy code, which need to modify and update the original buggy code. 
In contrast, our tool injects missing values to run partial code, we keep the original code untouched without changing any original code elements. 
Due to different goals and ways of generating code, APR methods are not applicable to our partial code execution task. 
}

To evaluate the effectiveness of our \toolName, we used the same dataset from Lexecutor containing two sets of code snippets: functions extracted from popular open-source projects and code snippets extracted from Stack Overflow posts. 
\xzp{Our results indicate that the \toolName enables the execution of 72.7\% and 83.3\% of all lines in the open-source code and Stack Overflow snippets, respectively, outperforming Lexecutor by 37.9\% and 33.5\%.}
Souza et al.~\cite{Souza2023LExecutorLE} first propose the task of partial code execution, and they use Lexecutor to find
the semantics-changing commits.
In this paper, we attempt to validate the practical usage of \toolName on a dynamic analysis task: runtime type error detection. 
\xzp{Specifically, by running \toolName on partial code fragment, our framework successfully detected 18 type error issues from eight popular Python GitHub repositories and 33 type error issues from Stack Overflow posts.}
In summary, this paper contributes the following:

\begin{itemize}[leftmargin=*]
    \item 
    We design and implement a framework, named \toolName, to engage the Code Llama model within LITL (LLM-in-the-loop) to guide the partial code execution. Our fine-tuned Code Llama model performs similarly to the close-source, commercial GPT-3.5 model in the task of guiding partial code execution. The richer complimentary dummy types help \toolName to guide more partial code execution.
    
    \item We extensively evaluate \toolName on both functions extracted from popular open-source projects and code snippets extracted from Stack Overflow posts. 
    The evaluation results show that \toolName can significantly outperform Souza et al~\cite{Souza2023LExecutorLE}'s method in both datasets \xzp{(37.9\% code coverage and 62.6\% fully executed rate improvement on the Open-source projects dataset, 33.5\% code coverage and 57.7\% fully executed rate improvement on Open-source projects dataset)}, achieving the state-of-the-art performance. 
    
    \item We validate \toolName with a practical dynamic analysis application: runtime type error detection. 
    \xzp{From eight popular Python GitHub repositories and 43 Stack Overflow posts, we successfully detected 18 and 33 type error issues, respectively.}
    To the best of our knowledge, our work is the first attempt to identify type errors at runtime, our tool can expose the runtime type error before compiling or running the entire software project, illustrating the effectiveness of our approach in practice.
\end{itemize}

\section{Motivation}
\label{sec:moti}
\begin{figure}[htbp]
    \centering
    \includegraphics{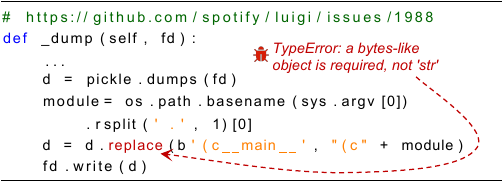}
    \caption{A Type Error Detected From Partial Code}
    \label{fig:moti}
\end{figure}

\xzp{The ability to execute partial code is essential for various dynamic analysis applications. 
We demonstrate a motivating example of checking runtime type errors using our approach, however, we argue that our approach is not limited to this particular application. 
It can be used to incorporate dynamic analysis tools to support a wide range of applications, for example, detecting security vulnerabilities via taint analysis. 
Better combining our tool with advanced dynamic checking techniques is an interesting future direction, but it is beyond the scope of our current research. 
} \\
\noindent\textbf{Motivating Example.} 
Python is one of the most popular programming languages nowadays. However, due to its dynamic type characteristics, variable types are determined and validated at runtime rather than compile time. 
Developers often suffer from runtime type errors when performing operations on inconsistent types of variables. 
Although Python static checkers (e.g., Pyre~\cite{Pyre} and mypy~\cite{mypy}) are designed to detect such type inconsistencies, however, they primarily rely on manually written type annotations which are unavailable most of the time.  As a result, Python type errors are often hard to detect unless they are exposed at runtime. 
Figure~\ref{fig:moti} demonstrates an example of Python type error in \textit{Luigi} project. 
Specifically, the method \texttt{replace} expects to be passed with two variables of the same type, in this case, both should be \texttt{bytes} objects. 
However, the developer wrongly passed a \texttt{string} object and thus introduced a type error. 
Due to complex internal dependencies, such type errors are difficult to trigger or reach out until bugs are eventually exposed. 
We manually checked the development history of the \textit{Luigi} project, this runtime type error has existed for over two years until finally exposed by a bug issue report. 
During this time, any code refactorings associated with this buggy method could be influenced, posing significant risks to software quality and maintenance. 
It is thus beneficial to have a tool that can discover such type errors without worrying about complex code dependencies or writing extensive test cases.

\noindent\textbf{\toolName Usage Scenarios.}
\toolName successfully detected this runtime type error without building/running the whole project. 
Based on the code snippet context, our framework correctly injects a \texttt{bytes} value object for the variable \texttt{d} and a \texttt{string} value object for the variable \texttt{module}, which successfully triggered the same runtime error reported by the bug issue report. 
Our framework can help developers expose this bug in an early stage (e.g., check-in time) and reduce the risks of introducing any unwanted problems or negative impacts. 
Suppose the developer who adopts our \toolName during his/her development, when code change happens, our tool can be performed on the newly updated partial code snippets for checking runtime type errors and discovering potential type errors just-in-time. 
It is worth mentioning that the usage scenario of our \toolName is not limited to runtime type error detection, our framework can be further extended to enable different dynamic analysis applications (e.g., assertion violation, taint analysis). 
In this work, we use runtime type error detection as a preliminary study to validate the practical usage of our framework. 

\section{Our Approach}
\label{sec:approach}
\begin{figure}[htbp] 
    \centering 
    \includegraphics[width=0.5\textwidth]{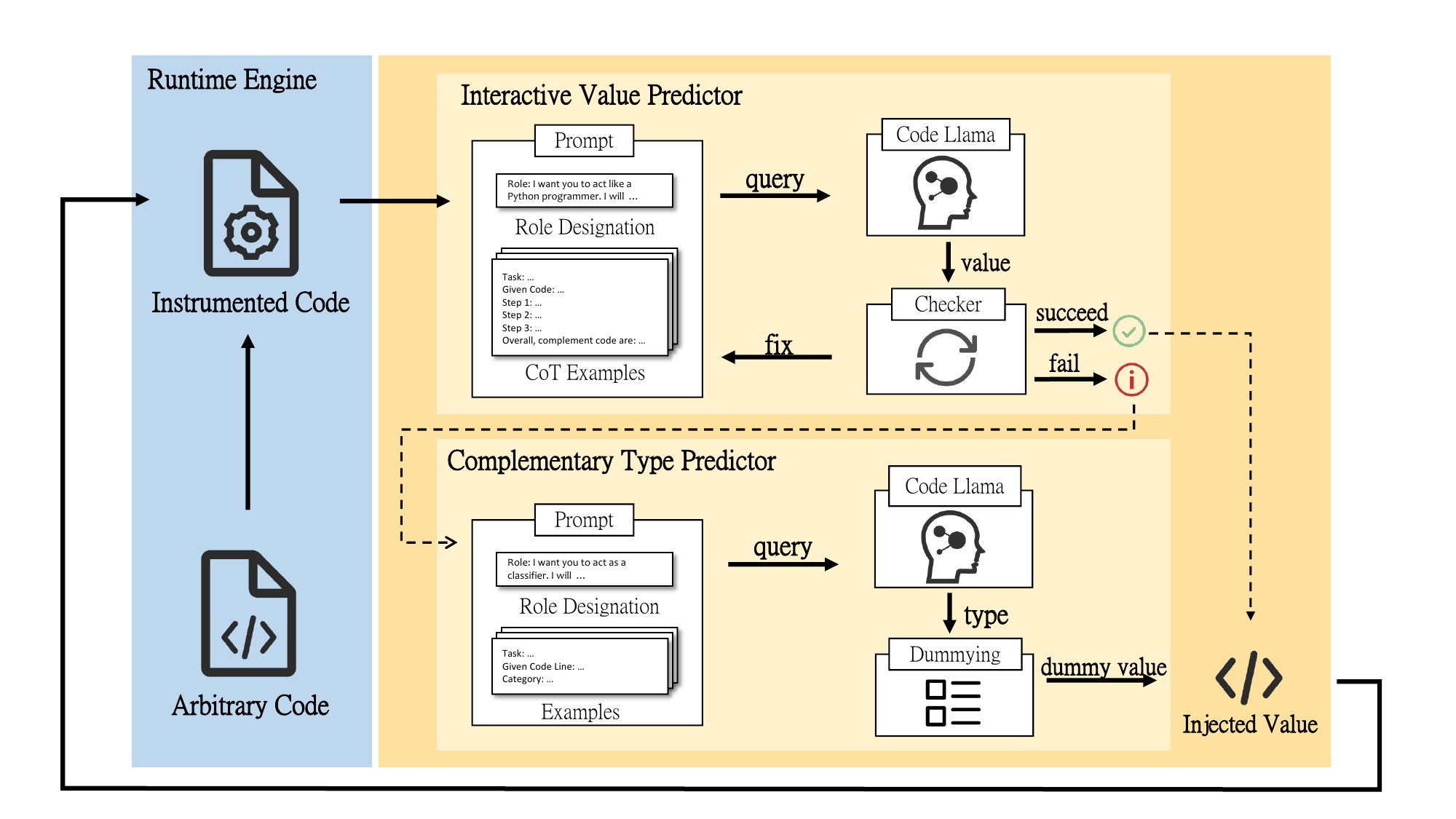} 
    \caption{The Overview of \toolName} 
    \label{fig:overview} 
\end{figure}

In this work, we design and implement an LLM-based framework, \toolName, to interactively make predictions and execute non-executable code snippets. 
\toolName includes three key components: the runtime engine, the interactive value predictor, and the complementary type predictor.
As shown in Fig.~\ref{fig:overview}, for a given non-executable arbitrary code snippet, the runtime engine first instruments arbitrary code snippets with execution hooks, and then executes the partial code and catches any exception that might be thrown when undefined code elements (e.g., variable, attribute) are met.
The raised exception will trigger the execution hooks, which send the undefined element and its contextual information to the interactive value predictor for inferencing the valid values for the undefined element. 
The execution hooks inject the inference values to the undefined element and guide code execution.
The interactive value predictor adaptively regenerates the likely values for the undefined elements and checks if these values can be executed by the runtime engine successfully.  
In certain cases, LLMs may fail to generate valid values even after multiple interactions, leading to the activation of the complementary type predictor. 
It queries the LLMs to predict the type of the undefined element and returns a pre-defined dummy value to the runtime engine. Details of each component are as follows.

\subsection{Runtime Engine}
The goal of the runtime engine is to catch the exception during partial code execution, query the interactive value generator, and inject the replied value to guide code execution.

The runtime engine initially instruments the arbitrary code with execution hooks from Lexecutor. It first visits the abstract syntax tree (AST) of the code and detects three types of AST nodes: variable reads, attribute reads, and calls of functions and methods. 
Then It instruments the detected three kinds of code by wrapping them with execution hooks.
The original code and instrumented code of each kind are illustrated in Table~\ref{tab:Execution Hooks}.
The iids refer to the instrument IDs, and the functions \_n\_, \_a\_, and \_c\_ are execution hooks for variable reads, attribute reads, and calls of functions and methods, respectively.

For the variable reads, the instrumented code calls the execution hook \_n\_, passing the name of the variable and a lambda
function that tries to read the value of the variable. \_n\_ then returns the value from the lambda function.
For the attribute reads, the instrumented code calls the execution hook \_a\_, passing the base object that has been assigned by \_n\_ and the name of the attribute. \_a\_ returns the value of the passing attribute of the base object.
For the calls of functions and methods, the instrumented code calls the execution hook \_c\_, passing the callee function. Then the execution hook invokes the callee function and returns the result of it.
During the execution of each hook, if it triggers some exception like \textit{NameError}, \textit{AttributeError}, the hook will query the interactive value predictor for a possible value. The query message of execution hooks combines the name of the code element, the kind of code element, the code line of the code element in the original code snippet located by the instrument IDs, and the error message during the execution.


\begin{table*}[htbp]
\setlength{\abovecaptionskip}{0cm}
  \centering
  \caption{Execution Hooks}
  \label{tab:Execution Hooks}
  \begin{tabular}{ccc}
    \toprule
    Ast Nodes & Original Code & Instrumented Code \\
    \midrule
    Variable Reads & Var1 = Var2 + 1 & Var1 = \_n\_(iid, "Var2", lamda: Var2) + 1 \\
    Attribute Reads & Opt1.Attr1 = Opt2.Attr2 & Opt1.Attr1 = \_a\_(iid, \_n\_(iid, "Opt2", lambda: Opt2), "Attr2") \\
    calls of functions \& methods & Var = Foo() & Var =  \_c\_(iid, \_n\_(535, "Foo", lambda: Foo))
 \\
    \bottomrule
  \end{tabular}
\end{table*}

\subsection{Interactive Value Predictor}
The interactive value predictor is to adaptively predict likely values of undefined code elements based on the contextual information and execution error message. 
The interactive value predictor combines a \textit{value generator} and a \textit{value checker}. 
The \textit{value generator} generates the definition or assignment of the queried undefined code element. 
The \textit{value checker} ensures the validity of the generated value by 
executing the definition or assignment and loading the value. 
If generated values are valid, the \textit{value checker} sends the loaded value to the runtime engine, and the code execution continues.
If generated values fail the validity checking, the \textit{value checker} will query back the \textit{value generator} again with a detailed execution error message. 
As a result, the interactive value predictor adaptively learns from the code execution results and progressively refines its predictions until executing the partial code successfully.  

\subsubsection{Interactive Value Generator}
\label{sec:prompt engineering}
The underlying approach of the value generator is prompt engineering. 
Role designation, few-shot learning and chain-of-thought reasoning are incorporated to construct LLM prompts. 
An example of the prompts is shown in Table~\ref{tab:interactive value predictor}. 


\textbf{Role Designation.}
In prompt engineering, role designation refers to designating LLMs with a specific role, providing them with a context that aids their understanding of the task context and leading to more accurate and relevant responses. 
In this study, since we aim to execute a Python code snippet, we designate the role of LLMs to act as a Python programmer. 
In addition, we also added output restriction and format restriction in the prompt. 
The prompt details are shown in Table~\ref{tab:interactive value predictor}.


\textbf{In-context Few-shot Learning.}
Few-shot learning is utilized to augment the context with a few examples of desired inputs and outputs. 
In this work, to select representative examples for few-shot learning, we invited three developers with at least five years of Python programming experience. 
Each of them was asked to fill in the likely values for the undefined code element based on contextual information and error execution message. 
After manually examining 30 arbitrary code snippets by each developer, the developers then meet and discuss the representation of the selection and refine the dataset until a consensus is reached. 
In total, we collect 6 examples as our representative dataset, with an example in Table~\ref{tab:interactive value predictor}.

\textbf{Chain-of-thought Reasoning.}
The in-context few-shot learning has provided LLMs with a few examples to learn the expected inputs and outputs, but the LLMs still lack the logical thinking to address the complicated task. 
We introduce the method of chain-of-thought to elicit the ability of LLMs' reasoning and logical thinking for this study. 
It endows the LLMs to split a complex task into several relatively simple steps and generate a series of intermediate outputs that lead to a reasonable result. Following the previous studies~\cite{52081, Wei2022ChainOT, Trivedi2022InterleavingRW}, we design a three-step thinking process that leads to the prediction of likely values for the undefined code elements. 
An example of chain-of-thought reasoning is shown in Table ~\ref{tab:interactive value predictor}. 
In particular, given a Python code snippet (e.g., \texttt{filepath = self.path}) which is non-executable:
In step 1, LLMs are required to import necessary modules for the code snippet, in this case, the \texttt{os} module is imported which is relevant to \texttt{filepath} within the code snippet. 
In step 2, LLMs are required to define all the necessary class/method/variable undefined of the code snippet, in this case, the class \texttt{MyClass} is defined and instantiated. 
Two types of step 3 are designed for our task, regarding step 3 (assign), LLMs are required to learn from the runtime engine and infer the likely values for the undefined code elements.  
In particular, three types of information are sent to the interactive value predictor, namely the undefined code element and its type (in this case the undefined element is \texttt{path} and its type is \texttt{attribute}), and the error execution message (\texttt{Attribute Error:`self' objects has no attribute `path'} for this case), LLMs is required to inference the likely values for the missing \texttt{path} attribute, generating \texttt{self.path = os.path.abspath(\_\_file\_\_)}. 
Regarding step 3 (fix), LLMs are required to interactively fix the last round's predicted values based on the failed execution message (\texttt{NameError: name `path' is not defined}) and last step output non-executable code (e.g., \texttt{self.path = path}). 
Finally, LLMs are required to summarize the aforementioned steps as outputs as shown Example Output in Table~\ref{tab:interactive value predictor}. 


\textbf{Prompt construction.} 
We combine the aforementioned information, i.e., ($<$\textit{Role Designation}$>$ + 6 * ($<$\textit{Chain-of-thought reasoning with Example Input}$>$ + $<$\textit{Example output}$>$)), to make two types of input prompt. 
Particularly, step 3 (assign) was used to construct the \textit{initial assign prompt}, and step 3 (fix) was used to construct the \textit{interactive fix prompt}. 
Two types of input prompts are constructed for the interactive value predictor, i.e., the \textit{initial assign prompt} and \textit{interactive fix prompt}. 
The \textit{initial assign prompt} is used for initiating the interactive value predictor while the \textit{interactive fix prompt} is used to fix undefined errors from the last round's predicted values.
Due to the advantage of few-shot learning and chain-of-thought reasoning, the LLMs will consistently reply to a Python code to assign a target code element in the same format as our example output, which can be directly executed.


\begin{table}[htbp]
\setlength{\abovecaptionskip}{0cm}
  \centering
  \caption{The Example of Prompt Engineering}
  \label{tab:interactive value predictor}
  \begin{tabular}{p{2.5cm}|m{5.2cm}}
    \toprule
    Prompt Type & \makecell[c]{Instantiation} \\
    \bottomrule
    \toprule
    \multirow{6}{*}{Role Designation} & \textbf{Role:} \textit{I want you to act like a Python programmer. I will give you Python code and comments, you should write Python code according to the comments step by step.}\\

     & \textbf{Output Restriction:} \textit{Only give reply with Python code and Do not write explanations.}\\
      & \textbf{Format Restriction:} \textit{Your reply is limited to only one code block and should wrap with backticks.}\\
      \hline
      \multirowcell{30}{Chain-of-thought\\ Reasoning with \\Example Input} & \textbf{Task:} \textit{Complete and fix the given code to make it can be executed directly.}\\
      & \textbf{Given code:} \textit{Do not modify the given Python code or wrap it with function.}\\
       & \textcolor{tablered}{<filepath = self.path>}\\
       & \textbf{Step 1:} \textit{Import needed module.}\\
       & \textcolor{blue}{import} os\\
       & \textbf{Step 2:} \textit{Define all the needed classes, methods, or variables here in detail.}\\
       & \textcolor{blue}{class} MyClass():\\
       & \textcolor{blue}{\quad pass}\\
       & self = MyClass()\\
       & \textbf{Step 3 (assign template):} \textit{Define and assign} \textcolor{tablered}{$\langle$UNDEF ELE$\rangle$ $\langle$UNDEF ELE TYPE$\rangle$} \textit{to repair the error} \textcolor{tablered}{$\langle$ERR MSG$\rangle$}\\
        & \textbf{Step 3 (assign case):} \textit{Define and assign} \textcolor{tablered}{$\langle$\texttt{path}$\rangle$ $\langle$\texttt{attribute}$\rangle$} \textit{to repair the error} \textcolor{tablered}{$\langle$\texttt{Attribute Error: `self' object has no attribute `path' for this case}$\rangle$}\\
       & {self.path = os.path.abspath(\_\_file\_\_)}\\
      & \textbf{Step 3 (fix template):} \textit{Fix the} \textcolor{tablered}{$\langle$LAST STEP CODE$\rangle$} \textit{since the} \textcolor{tablered}{$\langle$FAIL EXEC RES$\rangle$
      }. \\
      & \textbf{Step 3 (fix case):} \textit{Fix the} \textcolor{tablered}{$\langle$\texttt{self.path = path}$\rangle$} \textit{since the} \textcolor{tablered}{$\langle$\texttt{NameError: name `path' is not defined}$\rangle$
      }. \\
      & {path = os.path.abspath(\_\_file\_\_)}\\
      & {self.path = path}\\
       \hline
       \multirow{7}{*}{Example Output} & \textbf{Overall, complement code are:}\\
       & \textcolor{blue}{import} os\\
       & \textcolor{blue}{class} MyClass():\\
       & \textcolor{blue}{\quad pass}\\
       & {self = MyClass()}\\
       & {path = os.path.abspath(\_\_file\_\_)}\\
      & {self.path = path}\\
    \bottomrule
  \end{tabular}
\end{table}

 \subsubsection{Execution Value Checker}
After the value generator inference the likely values for the undefined code elements, the value checker executes the code with predicted values and queries back the value generator if necessary. 
In particular, the value checker first identifies and attempts to import or install the required third-party module. Then it invokes the \texttt{exec} function to execute the replied code from the value generator.
if the code execution succeeds, the value checker loads the value of undefined code elements, and then it returns the loaded value and the replied code to the runtime engine.
Otherwise, the value checker queries back the value generator with the previous predicted code and the error execution messages for refining. 


\subsubsection{LITL (LLMs-In-The-Loop) Algorithm} 
The key idea of incorporating LLMs in this work is to put LLMs-in-the-loop, we designate the LLMs as expert developers capable of interactively learning from execution results and finally guiding the partial code execution tasks. 
We demonstrate the details of the LITL (LLMs-In-The-Loop) algorithms in Algorithm~\ref{algo:LITL Algorithm}.  
For a given arbitrary code snippet, LLMs interactively refine undefined code values and check these values by execution (lines 2 to 9). 
The \textit{initial assign prompt} is constructed and queried to the LLMs to generate values for undefined code elements (lines 2 to 3). 
The algorithm then attempts to execute the generated code and catches any runtime exception (lines 4 to 7). 
It returns the value (i.e., $R$ and $V$, the result code $R$ refers to the predicted definition or assignment of the unknown code element from the value generator, and Loaded value $V$ refers to the value of the unknown code element loaded from result code $R$ execution.) if the code snippet executes successfully and no exception occurs. 
Otherwise, the \textit{interactive fix prompt is constructed} and query back LLMs for refining its previous predictions (Line 8). 
Whenever Algorithm~\ref{algo:LITL Algorithm} reaches line 10, it has failed to generate valid values after \textit{t} times iteration. 
It then throws an \textit{InvalidValueError}, which triggers the complementary type predictor.

\begin{algorithm}[htbp]
\SetAlgoLined
\KwIn{Kind $k$, name $n$, contextual information $c$ of code and error message $e$}
\KwOut{Result code $R$ and Loaded value $V$}

\BlankLine
$Prompt \leftarrow InitializeAssignPrompt(k,n,c,e)$\\
\For{$i=1$ \KwTo $t$}{
$C \leftarrow$ query LLMs with $Prompt$\\
$V \leftarrow $ Execute $R + c$  and load value, or catch exception $e$\\
\If{no exception while executing and loading}{return $V$, $R$}\
$Prompt \leftarrow BuildFixPrompt(c, e, V)$\
}
\textbf{throw} \emph{InvalidValueError}\
\caption{LITL Algorithm}
\label{algo:LITL Algorithm}
\end{algorithm}

\subsection{Complementary Type Predictor}
The complementary type predictor acts as a backup component for the adaptive value predictor. 
In certain cases, arbitrary code elements exceed the maximum interactions and LLMs fail to predict the appropriate code value, then the complementary type predictor will be triggered to infer the type of target undefined code element and return a corresponding pre-defined dummy value. 
Different from Lexecutor using CodeT5 for training, \toolName fine-tunes Code Llama by using the same training set of Lexecutor. We extend the predefined data types and leverage prompt engineering for type prediction. The implementation details are as follows.

\subsubsection{Pre-defined Dummy Value}
A pre-defined dummy value is a placeholder or default value that is used in place of a real value when the real value is not yet known or not applicable. 
As shown in Table~\ref{tab:Pre-defined dummy value}, we reuse the built-in date type (including None, Boolean, Integer, Float, String, List, Tuple, Set, and Dictionary) and Function and Objects type (including Callable, Object, and Resource) defined by Lexecutor.
Since Python is the primary programming language for deep learning and data analysis, we extend their pre-defined abstraction classes with three popular data types from third-party libraries (Tensor, Array, and DataFrame). 
When generated data values are not within the aforementioned abstraction classes, a dummy object value is injected by our approach. 

\subsubsection{Prompt Engineering}
Similar to constructing a prompt for adaptive value predictor, we first design the role of LLMs as: 
\textit{I want you to act as a classifier, I will give a line of Python code and a <word> in the code. You will classify the <word> into a category from None, Boolean, Integer, Float, String, List, Tuple, Set, Dictionary, Tensor, Array, DataFrame, Callable, Resource, Object.} 
Then we restrict the output of LLMs: 
\textit{I want you to only reply with the classified category and nothing else. Do not explain the result.} 
Since the type predictor is a comparatively straightforward classification task, we randomly select an example of each abstraction class as a few-shot learning example. 
In the end, we input the undefined code element and its contextual information to LLMs for inference.

\begin{table}[htbp]
\setlength{\abovecaptionskip}{0cm}
  \centering
  \caption{Pre-defined Dummy Value}
  \label{tab:Pre-defined dummy value}
  \begin{tabular}{c|c|c}
    \toprule
     Type &  Abstract Class & Dummy Value\\
    \bottomrule
    \multirow{10}{*}{\makecell{Built-in\\Data Type}} & None & None\\
     & Boolean & True\\
     &Integer & 1\\
     & Float & 1.0\\
     & String & "a" \\
     & List &  [Dummy()] \\
     & Tuple &  (Dummy()) \\
     & Set &  set(Dummy())   \\
     & Dictionary &  {"a": Dummy()} \\
    \toprule
    \multirow{3}{*}{\makecell{Third-party\\Data Type}} & Tensor & torch.tensor([[1.0]])\\
     & Array & numpy.array([1])\\
     & DataFrame & pandas.DataFrame(\{"a": 1\})\\
     \toprule
     \multirow{3}{*}{\makecell{Functions\\ \&objects}} & Callable & DummyCall\\
      & Object & Dummy()\\
      & Resource & DummyResource()\\
      \toprule
      \multicolumn{2}{c|}{Others} & Dummy() \\

    \bottomrule
  \end{tabular}
\end{table}
\vspace{-10pt}

\subsection{Implementation}
\label{subsecimplementation}
For the LLMs, we use the Code Llama instruct model with 34B parameters~\cite{roziere2023code}, which is the state-of-the-art open-source LLMs for coding applications.
We fine-tuned it with the Lexecutor code element type dataset~\cite{Souza2023LExecutorLE} to perform the missing type prediction. 
We employed the Parameter-Efficient Fine-Tuning strategy (PEFT)\cite{liu2022few}, and Low-Rank Adaptation (LoRA)\cite{hu2021lora} to accelerate the fine-tuning process. 
Based on 4*A800 Gpus, the fine-tuning process followed these hyperparameters: learning rate of $3e^{-4}$, batch size of 128, 2 epochs, and a warmup ratio of 100. 
After fine-tuning, the fine-tuned model is then used as our interactive value predictor and complementary type predictor by further leveraging prompt engineering and chain-of-thought reasoning techniques. 

We set the number of few-shot learning examples for the adaptive value predictor and complementary type predictor as 6 and 15, respectively. The 6 examples in the adaptive value predictor include 2 examples for each kind of undefined code element and 15 examples in the complementary type predictor include an example for each kind of abstract class. According to a small-scale pilot study, we set the maximum threshold in the adaptive value predictor as 5.

\section{Evaluation}
\label{sec:eval}


\subsection{Experimental Setup}
\textbf{Dataset.} 
Following the experiment set in Lexecutor, we reused the same datasets which included two sets of code snippets: functions extracted from popular open-source projects and code snippets extracted from Stack Overflow posts.
To avoid potential bias, as the raw data set is not provided, we carefully followed the data collection steps in Lexecutor.
Particularly, we initially extracted all the functions from five popular projects evaluated in Lexecutor, and we randomly selected 200 samples from each project. 
As listed in Table~\ref{tab:dataset}, the dataset is composed of 1,000 randomly selected functions, that amount to 7,225 non-empty, non-comment lines of code. 
To build the Stack Overflow code snippets dataset, we search for questions with the tag Python, then we randomly select an answer and extract the code in the top 1,000 votes questions. After removing the code snippets with invalid syntax, we collected 586 code snippets involving 4,540 non-empty, non-comment lines of code. The detail of the Stack Overflow code snippets dataset is shown in Table~\ref{tab:dataset}.

\begin{table}[htbp]
\setlength{\abovecaptionskip}{0cm}
  \centering
  \caption{Detail of Experiment Datasets}
  \label{tab:dataset}
  \begin{tabular}{c|c|cc}
    \toprule
    \multicolumn{2}{c|}{Dataset} & Count & Loc \\
    \midrule
     \multirow{4}{*}{\makecell[c]{Open-source\\ Projects Functions}}& Black & 200 &2,162 \\
     & Flask & 200 & 1,100\\
     & Pandas & 200 &1,438\\
     & Scrapy & 200 &1,150\\
     & TensorFlow & 200 &1,375\\
     \midrule
     \multicolumn{2}{c|}{Stack Overflow Code Snippets} & 586 &4,540\\
    \bottomrule
  \end{tabular}
\end{table}

\noindent\textbf{Baseline.} We set up Lexecutor~\cite{Souza2023LExecutorLE} which achieves states-of-art performance in partial code execution guiding. 
Lexecutor fine-tuned the pre-trained models (i.e., CodeT5~\cite{Wang2021CodeT5IU} and CodeBert~\cite{feng2020codebert}) with collected $<$code, type$>$ tuple in the training phase. During the execution phase, the runtime engine inputs the CodeT5 model with code and injects a dummy value according to the predicted type. 
Following the instructions in the replication package, we first collected training and validation sets from five popular open-source projects and fine-tuned CodeT5 with the set hyperparameters same as Lexecutor (\xzp{denoted as \textbf{Lexecutor-CodeT5}}), i.e., learning rate, epochs, and batch size. We achieved 79.1\%, 86.9\%, and 90.2\% accuracy of the top-1,3,5 predictions, respectively. 
The evaluation results of it closely match the accuracy they reported, we thus are confident with our replication process for Lexeuctor. 
\xzp{
Since \toolName is based on Code Llama, to conduct a more fair comparison with Lexecutor, we replaced the CodeT5 of Lexecutor with Code LLama, denoted as \textbf{Lexecutor-CodeLlama}. 
Following the methodology described in Sec~\ref{subsecimplementation}, we fine-tuned the Code Llama model using the same strategy and hyperparameters as \toolName. 
we also replaced Code Llama with ChatGPT~\cite{chatgpt} within our framework as a baseline, denoted as \toolName-GPT-3.5, which is used without any fine-tuning. 
}

\noindent\textbf{Metrics.} We evaluate the ability of \toolName to guide code execution in average code coverage and fully executed rate. The \textbf{Code Coverage} refers to the ratio of the number of executed lines of code to the total number of lines of code in the program. If the entire line has been executed without crashing, we label this line as ``covered''.
\xzp{\textbf{The Branch Coverage} refers to the ratio of the number of executed branches to the total number of branches in the program.}
The \textbf{Fully Executed Rate} \xzp{measures how many of all code snippets we achieve 100\% line coverage. 
} 
\xzp{The temperature hyperparameter of pre-trained generative models (including both CodeT5 and CodeLlama) controls the randomness of generated outputs. 
To help code execution cover as many lines as possible, we set the temperature hyperparameter to 0.8. This introduces more randomness and diversity in the generated outputs, allowing for a wider range of possible responses. 
We calculate the above metrics by combining results from five independent executions.}
The higher the metrics score, the better the approach can guide the incomplete code execution.

\subsection{RQ1: Effectiveness of \toolName}
To measure the effectiveness of \toolName in covering and successfully executing non-executable code, we evaluated our approach on datasets constructed of popular open-source project functions and Stack Overflow code snippets. 
The evaluation results are shown in Table ~\ref{tab:Overall effectiveness evaluation}. 
\xzp{
Lexecutor-CodeLlama has its advantage over Lexecutor-CodeT5, this is reasonable because Code Llama is a more powerful LLM which is 523 times larger than CodeT5.    
The performance of \toolName is significantly better than Lexecutor-based models (i.e., Lexecutor-CodeT5 and Lexecutor-CodeLlama) in both open-source project functions and Stack Overflow code snippets in terms of all metrics. 
We attribute this to the ability of \toolName for interactive learning by refining its predictions from execution results. In addition, our extension of pre-defined dummy values also contributed to better results.
}
By comparing \toolName with \toolName-GPT-3.5, they achieve a very close performance on both datasets, suggesting the generalizability of our framework for incorporating different LLMs. 
Compared with ChatGPT which contains 175B parameters, Code Llama is much smaller with only 34B parameters. 
Nonetheless, \toolName can achieve a comparable or even better performance than GPT-3.5 after fine-tuning, verifying the effectiveness of the fine-tuning process.
\begin{table}[htbp]
\setlength{\abovecaptionskip}{0cm}
  \begin{center}
  \xzp{
  \caption{Overall Effectiveness Evaluation}
  \label{tab:Overall effectiveness evaluation}
  \resizebox{0.48\textwidth}{!}{
  \begin{tabular}{cccc}
    \toprule
    Approach& Metrics & \makecell[c]{Open-source\\ Projects Functions} & \makecell[c]{Stack Overflow\\ Code Snippets} \\
        \midrule
    \multirow{3}{*}{Lexecutor-CodeT5}& Code coverage & 0.527& 0.624\\
     &Branch Coverage &0.312 &0.474\\
     &Fully Executed Rate& 0.342 & 0.478 \\
             \midrule
             \multirow{3}{*}{Lexecutor-CodeLlama}& Code coverage & 0.542 & 0.641\\
     &Branch Coverage & 0.371 & 0.511\\
     &Fully Executed Rate& 0.373 & 0.502\\
             \midrule
             \multirow{3}{*}{\toolName-GPT-3.5}& Code Coverage &0.730 & 0.819\\
             &Branch Coverage &0.644 &0.789\\
 &Fully Executed Rate&0.556 & 0.746\\
 \midrule
         \multirow{3}{*}{\toolName}& Code Coverage & 0.727 & 0.833\\
         &Branch Coverage & 0.643 &0.795\\
 &Fully Executed Rate& 0.556 & 0.754\\
    \bottomrule
  \end{tabular}
  }}
  \end{center}
\end{table}

\begin{figure}[htbp]
  \centering
  \includegraphics[width=0.49\textwidth]{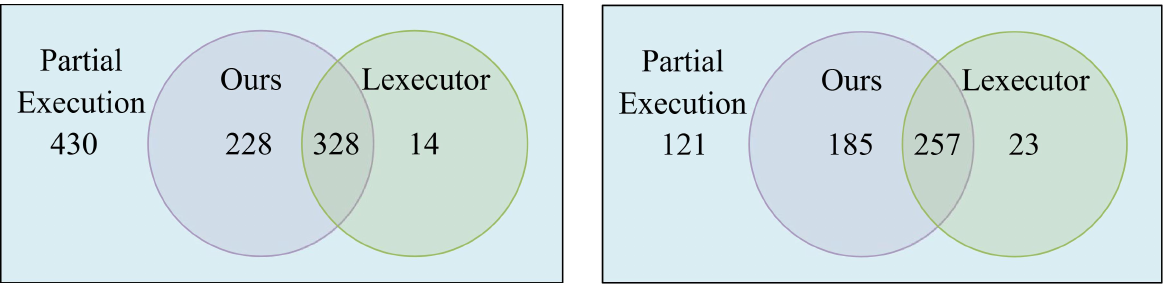}
  \caption{Venn Graph for Fully Executed Code Snippets Reported by Our Approach and Lexecutor, Open-Source Project Functions (left) and Stack Overflow Code Snippets (right)}
  \label{fig:RQ1VEEN}
\end{figure}

Fig.~\ref{fig:RQ1VEEN} demonstrates the Venn diagrams of fully executed code snippets reported by our approach and Lexecutor. 
We can find that, \textbf{most code snippets covered by Lexecutor are also covered by our approach, while our approach can cover far more cases Lexecutor can not handle.} 
For example, 228 project functions (40\%) and 185 Stack Overflow code snippets (39.8\%) are successfully executed by our approach but failed to be handled by Lexecutor. 
While only 14 project functions (2.5\%) and 23 Stack Overflow code snippets (4.9\%) are reported by Lexecutor but missed by ours. 
This further justifies the superiority of our proposed \toolName. 
We also observed that several cases can not covered by our approach and/or Lexecutor, We detailed discussed why we work and why we fail in Section~\ref{sec:discussion}.

\vspace{-10pt}


\subsection{RQ2: Component Analysis}
The performance of \toolName mainly relies on two components: the interactive value predictor and the complementary type predictor.
We evaluate the performance of each component respectively. 
In particular, we compare \toolName with two incomplete versions:
\begin{itemize}[leftmargin=*]
    \item Interactive value predictor. 
    In this version, we only keep the interactive value predictor and remove the complementary type predictor, the values are directly injected from the interactive value predictor. 
    \item Complementary type predictor. In this version, we only use the complementary type predictor to infer the type of queried element and inject the pre-defined value.
\end{itemize}

\vspace{-10pt}
\begin{table}[htbp]
\setlength{\abovecaptionskip}{0cm}
  \begin{center}
  \xzp{
  \caption{Effectiveness of two components in \toolName}
  \label{tab:Ablation Study}
  \resizebox{0.5\textwidth}{!}{
  \begin{tabular}{cccc}
    \toprule
    Approach& Metrics & \makecell[c]{Open-source\\ Projects Functions} & \makecell[c]{Stack Overflow\\ Code Snippets} \\
        \midrule
    \multirow{3}{*}{\makecell{Interactive Value\\Predictor}}& Code coverage & 0.636 &  0.744\\
    &Branch Coverage & 0.552 & 0.704\\
     &Fully Executed Rate& 0.477  & 0.679\\
     \midrule
     \multirow{3}{*}{\makecell{ Complementary Type\\Predictor}}& Code Coverage & 0.558 & 0.656 \\
     &Branch Coverage &0.390 & 0.521\\
    &Fully Executed Rate& 0.382& 0.514\\
   \midrule
  \multirow{3}{*}{\toolName}& Code Coverage &  0.727 & 0.833\\
  &Branch Coverage & 0.643 &0.795\\
    &Fully Executed Rate&0.556  & 0.754 \\
    \bottomrule
  \end{tabular}
  }}
  \end{center}
\end{table}
\vspace{-10pt}

The results are shown in Table~\ref{tab:Ablation Study}. From the tables, several points stand out: 
(i) No matter which component we removed, it reduces the performance of our approach in guiding partial code execution. 
This verifies the importance and usefulness of our interactive value predictor and complementary type predictor.
(ii) The interactive value predictor and complementary type predictor can complement and enhance the performance of each other. 
For example, the interactive value predictor performs better on Stack Overflow code snippets while the complementary type predictor achieves better performance on open-source project functions. 
Although both components use our fine-tuned Code Llama model for inference, the interactive value predictor focuses on generating likely values for undefined code elements, while the complementary type predictor focuses on predicting the types, the different learning objectives of these two sub-components make them a suitable pair to enhance each other's capabilities. 
As a result, after combining these two modules, 
the performance of \toolName is significantly boosted and achieved state-of-the-art performance. 



\subsection{RQ3: Sensitivity Analysis}
The interaction learning and chain-of-thought reasoning are the core mechanisms of our approach. 
To explore the effectiveness of the above mechanisms, we construct a sensitivity analysis. 
To demonstrate the effectiveness of interactive learning from code execution results, we evaluate the performance of \toolName after each iteration, to demonstrate the effectiveness of chain-of-thought reasoning, we evaluate \toolName using the prompt without chain-of-thought settings. 
It is worth mentioning that to better present the performance contributed by each mechanism alone, we drop the complementary type predictor for this RQ setting. 

\begin{figure}[htbp]
\centering
\includegraphics[width=0.49\textwidth]{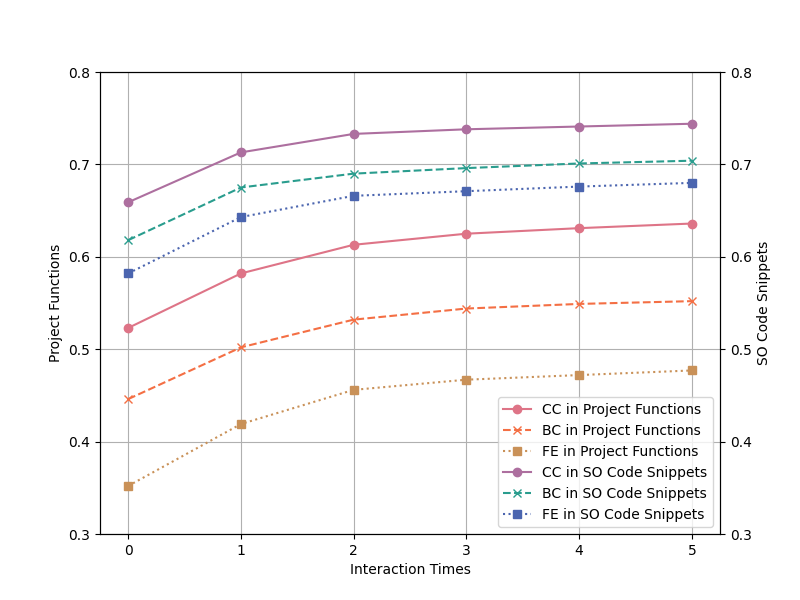}
\caption{Performance of Interactive Value Predictor}
\label{fig:RQ3}
\end{figure}

Fig.~\ref{fig:RQ3} illustrates the performance of our approach under different interaction times. 
We found that the performance of our approach rapidly increased after the initial two interactions. 
Regarding simple problems such as introducing unknown variables or parameters, can be easily resolved after one or two iterations of interactively learning. 
This further confirms the self-guided interactive learning ability of our fine-tuning Code Llama model. 
Then the improvement ratio slows down after 3 interactions, the reason for this can be the insufficient code context information and/or unclear code execution error message, which indicates that the fine-tuning Code Llama model is helpful but not a 'silver bullet' for value prediction. 
The 5 interaction times we used in our settings are reasonable for achieving optimal results.

Table~\ref{tab:chain-of-thought reasoning} illustrated the performance of our approach (drop complementary type predictor) with and without chain-of-thought reasoning.
\xzp{The code coverage, branch coverage and fully executed rate decreased by 24.7\%, 61.4\% and 45.9\% on the open-source project functions dataset and those decreased by 15.3\%, 31.1\% and 27.5\% on the Stack Overflow code snippets dataset. }
We found that, without the chain-of-thought reasoning, the logic reasoning capability of our approach drops significantly, for example, the fine-tuning Code Llama model often uses a module before importing it (as defined in step 1), query undefined classes or variables (as defined in step 2), this further justifies the effectiveness of chain-of-thought reasoning for prompt engineering. 

\begin{table}[htbp]
\setlength{\abovecaptionskip}{0cm}
  \begin{center}
  \xzp{
  \caption{Effectiveness of Chain-of-thought Reasoning}
  \label{tab:chain-of-thought reasoning}
  \resizebox{0.48\textwidth}{!}{
  \begin{tabular}{cccc}
    \toprule
    Approach&Metrics & \makecell[c]{Open-source\\ Projects Functions} & \makecell[c]{Stack Overflow\\ Code Snippets} \\
    \midrule
    \multirow{3}{*}{\makecell{Interactive Value\\Predictor w/o CoT}}& Code Coverage &  0.479& 0.630\\
    &Branch Coverage & 0.213 & 0.485\\
    &Fully Executed Rate& 0.258 & 0.493\\
     \midrule
     \multirow{3}{*}{\makecell{Interactive Value\\Predictor}}& Code Coverage & 0.636 & 0.744\\
     &Branch Coverage &0.552 & 0.704\\
     &Fully Executed Rate& 0.477 & 0.680\\
    \bottomrule
  \end{tabular}
  }}
  \end{center}
\end{table}


\subsection{\xzp{RQ4: Time Cost Analysis}}
\xzp{In this RQ, to evaluate the efficiency of \toolName, we conduct the time cost analysis regarding two aspects: (i) We compare the time cost of each prediction taken by \toolName and Lexecutor; (ii) We compare the performance of Lexecutor and \toolName by allocating them with the same time budget. 
}


\xzp{
For the first aspect of time cost analysis, the prediction time of Lexecutor costs from 0.18s to 0.48s, while SelfPiCo takes 6.25s to perform a single prediction. We found that the time cost of \toolName is largely due to the interactive learning process (i.e., 2.28s for a single round of interactive learning in open-source project functions). Besides, the model in Lexecutor only needs to output a predicted type, while \toolName must generate a concrete code snippet. The time cost of \toolName can be reduced with parallelization and more advanced hardware. Moreover, we argue that \toolName is a general framework that can easily be incorporated with other smaller-size LLMs, further decreasing the time costs. }

\xzp{Regarding the second part analysis, we equitably compared the performance of Lexecutor by allocating it with the same time spent by \toolName. Specifically, we set different temperature values to run multiple-rounds of execution until reaching the time limit. The experimental results remained the same (e.g., 0.527 code coverage on the Open-source dataset and 0.624 code coverage on the StackOverflow dataset) between the basic five-round executions in the previous experiment and subsequent multiple-round executions, suggesting allocating extra time to Lexecutor can’t bring performance gains. }

\subsection{Result Discussion}
\label{sec:discussion}

\noindent\textbf{Why \toolName works.} 
As shown in Fig.~\ref{fig:RQ1VEEN}, there are 86 functions and 125 code snippets that can be fully executed by our approach but failed by Lexecutor. 
We summarize three advantages of our approach over Lexecutor, including valid value injection, accurate type prediction, and comprehensive data types. 

In particular, compared with Lexecutor: \textbf{First, we generate more accurate values} for the undefined code elements by self-guided interactive learning. 
The Lexecutor uses the pre-defined dummy values to fill the code, to compare, 
the interactive learning of \toolName can dynamically assign and refine the required values based on code context and execution results.
As the Listing~\ref{listing:introlist} shows, Lexecutor successfully predicts the correct type (i.e., \textit{Callable}) for the undefined \texttt{filter\_cached}, but the pre-defined value \textit{DummyObject} conflicts the expected return value. 
According to the error message, our approach guides the model to assign the value as a tuple, which can be unpacked into two values and successfully address the issue. 
\textbf{Secondly, we predict more accurate types with the fine-tuned Code Llama model.}
Compared with CodeT5, LLMs are trained on ultra-large-scale datasets and exhibit promising performance in code understanding and logical reasoning, which have achieved great accuracy on type prediction~\cite{Ding2023TRACEDEP}. 
Listing~\ref{listing:discussionexample} - Ex.1 illustrates an example where the queried code element \texttt{declarations} is a list, but Lexecutor predicts it as \textit{DummyObject}, resulting in a type error. 
In contrast, our approach accurately identifies the type of \texttt{declarations} and assigns it with a tuple, enabling successful value retrieval later on. 
\textbf{Thirdly, we apply more comprehensive data types with third-party libraries.} 
In Listing~\ref{listing:discussionexample} - Ex.2, the unknown code element \texttt{df} is actually a \texttt{DataFrame} provided by the \texttt{pandas} module. 
None of the pre-defined dummy values in Lexecutor could be injected appropriately. 
However, our approach overcomes this limitation by importing the \texttt{pandas} module and defining the \texttt{DataFrame} as an extended third-party data type, allowing code execution successfully continues.

\begin{lstlisting}[caption=Successful \& Failed Cases of \toolName, label=listing:discussionexample, escapeinside=||]
# Ex1: Accurate type prediction
# Original Code: black/src/black/concurrency.py:
for prop, value in |\underline{declarations}|:
    prop = prop.lower()
    value = value.lower()
# Lexecutor Injection:
declarations = DummyObject
|\color{red}TypeError: cannot unpack non-iterable DummyObject object|
# SelfPiCo Injection:
declarations = [("color", "red")]
--------------------------------------------------
# Ex2: Comprehensive data types
# Original Code: pandas/tests/groupby/transform/test_transform.py:
expected = df[-|\underline{df}|.a.isin(drop_idx.index)]
# Lexecutor Injection:
df = DummyObject
|\color{red}TypeError: bad operand type for unary -: 'DummyObject'|
# SelfPiCo Injection:
df = pd.DataFrame({'a': 1})
--------------------------------------------------
# Ex3:Insufficient code Instrumentation & Inadequate contextual information
# Original Code: black/src/black/trans.py:
LL = |\underline{line}|.leaves
|…|
if LL[comma_idx].type == token.COMMA:
# SelfPiCo Injection:
class Line:
    def __init__(self, leaves):
        self.leaves = leaves
line = Line([])
|\color{red}IndexError: list index out of range|

\end{lstlisting}


\noindent\textbf{Limitations of \toolName.} We also investigate why our \toolName fails to execute certain partial codes, two main reasons are identified, as shown in Listing~\ref{listing:discussionexample} - Ex.3: 
\textbf{Inadequate contextual information.} In our approach, we only input the line where the undefined code element is located as the contextual information. 
The generated value may satisfy the current line execution requirement but conflict with the subsequent code. 
For example in Ex.3, without the following contextual information about \texttt{LL}, \toolName assigns an empty list for it. 
\textbf{Insufficient code instrumentation.} In the code instrumentation phase, we utilize the Lexecutor and instrument three kinds of execution hooks: variable reads, attribute reads, and calls of functions. 
However, these hooks are insufficient and miss some important operations, such as indexing, and binary operation. For example, the empty list assignment to \texttt{LL} results in an \textit{IndexError} when indexing operation \texttt{LL[comma\_idx]} is performed. However, none of the exception hooks can catch the \textit{IndexError}, leading to the termination of code execution.

\section{Practical Applications}
\label{sec:Applications}
\subsection{Runtime Type Error Detection}
In this section, we apply our \toolName in a real dynamic program analysis task: runtime error detection. 
As discussed in the Motivation Section, Python runtime errors are often hard to discover and/or trigger until the bugs are eventually exposed. 
\xzp{To verify the practical usage of our framework, we apply \toolName in real GitHub projects and Stack Overflow posts to assess its effectiveness.}
Particularly, to collect the real type errors from GitHub, we first selected eight popular GitHub open-source projects that have more than 1000 stars (i.e., Pandas, Airflow, Luigi, Ansible, Core, Keras, Requests, and Salt). Then for each project, we searched the pull requests containing the keyword \textit{Type Error} to find those reporting and fixing type error issues. 
We excluded type errors whose messages included project-specific domain knowledge that could not be generated.  
Finally, we collected 42 type errors from the above eight projects for our evaluation. For each collected type error, we check out the code snippet that introduced type error as our partial code input.
\xzp{To collect real type errors from Stack Overflow posts, we randomly selected 200 posts with the keyword ``Type Error'' in the title or body. After filtering out code snippets containing invalid syntax, we collected 47 unique type errors for analysis.}
We arranged each partial code into a separate file and then leveraged \toolName to run the partial code to see if the target type error could be successfully triggered. 
If and only if our approach terminates at the same fault localization and reports the same error messages with the issues, we consider this type error as successfully detected.

\begin{figure}
    \centering
    \includegraphics{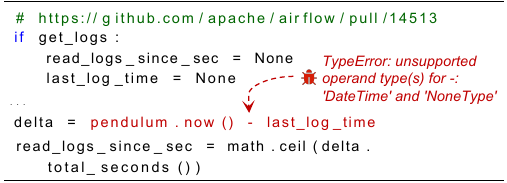}
    \caption{Type error detected by \toolName}
    \label{fig:case study}
    \vspace{-10pt}
\end{figure}

\xzp{\toolName successfully detected 18 Python type errors from 42 collected ones from the GitHub projects and 33 Python type errors from 47 collected ones from the Stack Overflow forum. Lexecutor-CodeLlama can only detect 8 and 21 Python type errors from GitHub projects and Stack Overflow forum, respectively.}
Figure~\ref{fig:case study} shows a detected type error. First, \toolName injects the value of \texttt{True} for \texttt{get\_logs}, and the variable \texttt{last\_log\_time} is assigned with \texttt{None}. 
When executing the buggy line, \toolName predicted the undefined variable \texttt{pendulum} as an object which includes a method \texttt{now} to return the current time. 
Then \toolName successfully detected the runtime type error: \textit{TypeError: unsupported operand type(s) for -: 'DateTime' and 'NoneType'}. 
The runtime type error detection shows the practical value of our tool to facilitate the dynamic program analysis of applications. However, there are still cases \toolName can not handle correctly, the failed cases primarily stem from cases where the partial codes are too complicated to perfectly handle, and/or insufficient context to infer the error triggered types (e.g., \texttt{last\_log\_time} may also be assigned with \texttt{DateTime} object). 
\xzp{Furthermore, to estimate the false positive rate of SelfPiCo, we randomly sampled 50 function bodies from the 8 open-source projects and 50 partial codes without error from the Stack Overflow forum. 
Following that, we run SelfPiCo to execute these 100 partial codes to see if any potential type errors will be triggered. 
The experimental results show that SelfPiCo reported 11 type errors for 100 partial codes, resulting in a false positive rate of 11\%, while Lexecutor-CodeLlama performed a 28\% false positive rate. The relatively low false positive ratio further confirms the practical usage of our approach. 
We manually checked false positive cases, these failed cases are primarily caused by imprecise value predictions (7 cases due to lack of code context, 4 cases due to complex variable value), it would be interesting to address these limitations in future work. 
} 

\subsection{\xzp{Discussion}}

\xzp{Unit test generation (UTG) technology is widely used to detect runtime errors, recent research also leveraged LLMs to generate unit test~\cite{lemieux2023codamosa, schafer2023empirical, yuan2023no}. 
For example, Schafer et al.~\cite{schafer2023empirical} introduced the LLM-based model TestPilot to generate tests by re-prompting the model with error messages, Yuan et al.~\cite{yuan2023no} proposed a ChatGPT-based model ChatTester to leverage ChatGPT to improve unit test generation. 
UTG methods differ from our research as follows: 
(i) \textbf{UTG methods are incapable of handling partial code.} 
UTG methods, such as TestPilot and ChatTester, require the method under test (i.e., focal method) can be invoked and executed properly. 
The underlying assumption is that focal methods should be complete and compilable, while either our open-source functions (uncompilable) or Stack Overflow code snippets (incomplete and uncompilable) fail to satisfy such conditions. 
In other words, the partial code can not be directly invoked and executed, making UTG tools unable to generate tests for them. 
For example, TestPilot and ChatTester can't generate unit tests for SO code snippets, because 75\% SO code snippets can not be executed. 
Moreover, SO code snippets are often code lines and lack method signatures, rendering UTG tools ineffective;
(ii) \toolName \textbf{can discover different runtime errors that UTG can not detect}. 
Based on the focal method, UTG tools (e.g., TestPilot and ChatTester) generate a unit test that invokes the target method with reasonable input parameter values and checks the output with corresponding assertions. 
UTG injects values only at well-defined interfaces, such as function entry points. 
While \toolName can inject valid runtime values in arbitrary points of the code during execution on-demand, enabling the discovery of bugs that are not triggered by changing input value changes.  
Such as the state-dependent issues and boundary condition problems, these bugs can't be essentially checked by UTG methods, \toolName can assist developers to find these hidden bugs in runtime. 
}

\section{THREATS TO VALIDITY}
\label{sec:threats}
In our experiments evaluating our model, threats to internal validity may arise from the randomness of LLMs generation, which may generate different results for different runs. It means LLMs may reply with different outputs based on the same input. To mitigate this threat, we calculated the metrics by combining results from five independent executions.

The main external threat to the validity of our work is the representative of the testing dataset selected to evaluate our approach. To
mitigate this threat, we followed the same strategy as the baseline method, representing an unbiased testing dataset for our study. 
\xzp{Moreover, Our practical evaluation is based on known issues confirmed or reported by developers, these data samples can be regarded as ground truth and we can easily measure the effectiveness of our \toolName on these samples. 
It would be interesting future research direction to use our tool to detect more runtime-type errors in the wild. }


\section{Related Work}
\label{sec:related work}

\textbf{Incompletion code execution.} micro-execution~\cite{Godefroid2014MicroE} builds a runtime Virtual Machine that allows for executing arbitrary x86 code by injecting binary values into memory on demand. \xzp{X-force~\cite{peng2014x} executes arbitrary binary code and fixes the invalid memory by setting the offending pointers to the allocated memory.} UC-KLEE~\cite{Ramos2015UnderConstrainedSE} extends the symbolic execution (KLEE) for an incompletion code snippet. J-Force~\cite{Kim2017JForceFE} forced to execute the uncovered path and inject the value candidates from data flow for missing objects. \xzp{JSForce~\cite{tang2018dual}, Dual-Force~\cite{hu2018jsforce} and Oyama et al.~\cite{oyama2023forced} explored the execution paths of arbitrary malware code by switching between different execution paths when encountering exceptions.}
LExecutor~\cite{Souza2023LExecutorLE} predicted the type of missing code element and injected the corresponding injected pre-defined dummy value. Our work fundamentally differs from the above approach by predicting the definition and assignment of missing code elements and injecting realistic value.

\noindent\textbf{Execution behaviour analysis}
Since several tasks require the behavior of code execution, some research focuses on predicting the behavior of code execution. Bieber et al. ~\cite{Bieber2020LearningTE} proposed an instruction pointer attention graph neural network (IPA-GNN) to infer the runtime value of each variable. Some research aimed to predict the type of dynamic Language~\cite{mir2022type4py, Peng2021StaticIM} or binaries~\cite{Pei2021StateFormerFT, Lehmann2022FindingTD}. TRACED~\cite{Ding2023TRACEDEP} fine-tuning the large language model by the execution trace of code and predicting the execution branch of code without execution. Moreover, Bieber et al.~\cite{Bieber2022StaticPO} predicted whether a program has runtime errors or an exception raised. The approach mentioned above illustrated the feasibility of predicting the runtime behavior of the program. Compared to all the above work, our approach not only predicts the runtime value of the code element but also practically executes the program.

\noindent\xzp{\textbf{Automated  Program Repair.} APR tries to modify a program to achieve successful compilation or execution, which overlaps to some extent with our partial code execution. A common approach to APR regards it as a code transformation task, which transforms the buggy program into a bug-fixing program~\cite{10.1145/3377811.3380345, 9401997, zhu2021syntax, ye2022selfapr}. Recent research has explored the potential of large language models for program repair~\cite{xia2023keep}. Several studies have demonstrated that LLMs display a basic level of program comprehension that can be for APR~\cite{jin2023inferfix, joshi2023repair, xia2023automated, deligiannis2023fixing}. Different from APR, our study focuses on partial code execution and will not modify the target program or its semantics.}

\noindent\textbf{Dynamic analysis for Python.} Dynamic analysis is crucial in program analysis, and there is some research work for Python dynamic analysis. For the runtime dynamic analysis, Xu et al.~\cite{Xu2016PythonPA} collected the execution trace of the Python program and leveraged the SMT solver to detect bugs. Chen et al~\cite{Chen2014DynamicSO} instrumented the bytecode of the Python program and executed instrumented bytecode to capture the data and control flow and sliced the file. SCALENE~\cite{Berger2020ScaleneSA} is a high-performance CPU, GPU, and memory profiler for Python, which monitors memory usage during Python program execution. DynaPyt~\cite{fse2022-DynaPyt} is a dynamic analysis framework that instruments the code with the analysis hooks and supports customized dynamic analysis tasks. Fuzzing technology is widely used for bug identifying during execution\cite{203944, Deng2022LargeLM, Wei2022FreeLF}. The above approaches need to execute the Python code, while our approach can support them to analyze non-executable Python code.
For compile-time dynamic analysis, angr~\cite{shoshitaishvili2016state} translated the binary code into an intermediate representation (IR) and performed symbolic execution. Triton is a dynamic analysis applied taint analysis and symbolic execution on the instrumented IR. Since the non-executable code can not be translated to valid IR, our approach can complement the code to be successfully compiled and executed.

\noindent\xzp{\textbf{Unit Test Generation.} UTG is widely used to detect errors dynamically, which can also detect the runtime time errors. Based on the target function, these tools produce a unit test that invokes the target function with reasonable input parameter values and checks the output with corresponding assertions~\cite{zhang2023repocoder, ma2020can}.  Recently there are also some tools leveraging LLMs to generate unit test~\cite{lemieux2023codamosa, schafer2023empirical, yuan2023no}. 
A key difference to our work is that UTG assumes that the target method is complete and compilable, which can be invoked and executed directly, while our \toolName aims to detect runtime type errors from un-executable partial code.}

\section{Conclusion and Future Work}
\label{sec:conclusion}

Aim to dynamically analyze arbitrary code, e.g., non-executable partial code snippets, we introduce \toolName leveraging the powerful learning capabilities of LLMs to interactively fill in the partial code to make it executable. 
The experiments demonstrate the effectiveness of our approach in guiding partial code execution.
\xzp{The exploratory study of \toolName on the practical usage shows that \toolName successfully detects 51 type errors, illustrating the usefulness of our approach in arbitrary code dynamic analysis.} \xzp{In this study, we only use error messages for applying LLMs, we will explore other domain information during code execution (such as partial AST and API sequence) in our future work.}


\section{Data Availability}
\label{sec:data}
Our replication package is available at ~\cite{replication_package}. 

\begin{acks}
This research is supported by the Starry Night Science Fund of Zhejiang University Shanghai Institute for Advanced Study, Grant No. SN-ZJU-SIAS-001. 
This research is partially supported by the Shanghai Sailing Program (23YF1446900) and the National Science Foundation of China (No. 62202341). 
This research is partially supported by the Ningbo Natural Science Foundation (No. 2023J292). 
This research was also supported by the advanced computing resources provided by the Supercomputing Center of Hangzhou City University. 
The authors would like to thank the reviewers for their insightful and constructive feedback.
\end{acks}

\balance
\bibliographystyle{ACM-Reference-Format}
\bibliography{sample-base}
\end{document}